\input{psfig.sty}
\documentclass[12pt,preprint]{aastex}
\usepackage{epsfig}

\shorttitle{The Afterglow of GRB\,050319}

\begin{document}

\title{Swift XRT Observations of the Afterglow of GRB\,050319}

\author{
Giancarlo~Cusumano\altaffilmark{1},
Vanessa~Mangano\altaffilmark{1},
Lorella~Angelini\altaffilmark{2,3},
Scott~Barthelmy\altaffilmark{2},
Andrew~P.~Beardmore\altaffilmark{4},
David~N.~Burrows\altaffilmark{5}, 
Sergio~Campana\altaffilmark{6},
John~K.~Cannizzo\altaffilmark{2},
Milvia~Capalbi\altaffilmark{7},
Guido~Chincarini\altaffilmark{6,8},
Neil~Gehrels\altaffilmark{2},
Paolo~Giommi\altaffilmark{7}, 
Michael~R.~Goad\altaffilmark{4},
Joanne~E.~Hill\altaffilmark{5},
Jamie~A.~Kennea\altaffilmark{5},
Shiho~Kobayashi\altaffilmark{5},
Valentina~La Parola\altaffilmark{1},
Daniele Malesani\altaffilmark{9},
Peter~M\'{e}sz\'{a}ros\altaffilmark{5},
Teresa~Mineo\altaffilmark{1},
Alberto~Moretti\altaffilmark{6},
John~A.~Nousek\altaffilmark{5}, 
Paul~T.~O'Brien\altaffilmark{4},
Julian~P.~Osborne\altaffilmark{4},
Claudio~Pagani\altaffilmark{5,6},
Kim~L.~Page\altaffilmark{4},
Matteo~Perri\altaffilmark{7},
Patrizia~Romano\altaffilmark{6},
Gianpiero~Tagliaferri\altaffilmark{6}, 
Bing~Zhang\altaffilmark{10}
}

\altaffiltext{1}{INAF- Istituto di Astrofisica Spaziale e Fisica Cosmica Sezione di Palermo,  
                 Via Ugo La Malfa 153, 90146 Palermo, Italy; {\it cusumano@pa.iasf.cnr.it}}
\altaffiltext{2}{NASA/Goddard Space Flight Center, Greenbelt, MD 20771}
\altaffiltext{3}{Johns Hopkins University}
\altaffiltext{4}{Department of Physics and Astronomy, University of Leicester, Leicester LE1 7RH, UK}
\altaffiltext{5}{Department of Astronomy \& Astrophysics, 525 Davey
  Lab., Pennsylvania State
University, University Park, PA 16802, USA}
\altaffiltext{6}{INAF -- Osservatorio Astronomico di Brera, Via Bianchi 46, 23807 Merate, Italy}
\altaffiltext{7}{ASI Science Data Center, via Galileo Galilei, 00044 Frascati, Italy}
\altaffiltext{8}{Universit\`a degli studi di Milano-Bicocca,
                 Dipartimento di Fisica, Piazza delle Scienze 3, I-20126 Milan, Italy}
\altaffiltext{9}{International School for Advanced Studies (SISSA-ISAS), via Beirut 2-4, I-34014 Trieste, Italy}
\altaffiltext{10}{Department of Physics, University of Nevada, Las Vegas, NV 89154-4002, USA}

\begin{abstract}
Swift discovered the high redshift GRB\,050319 
with the Burst Alert Telescope and began observing with its narrow
field instruments only 225 s after the burst onset.
The afterglow X-ray emission was monitored by the XRT up to 28 days 
after the 
burst. The light curve shows a decay with  
three different phases, each characterized by a 
distinct slope: an initial steep decay with a power law index 
of $\sim$ 5.5, a second phase characterized by a flat decay slope 
of $\sim$ 0.54, and a third phase with a decay slope of $\sim$ 1.14.
During the first phase the spectral energy distribution is softer than
in the following two phases
and the photon index is consistent with the GRB prompt spectrum.
The extrapolation of the BAT light curve to the XRT band suggests
that the initial fast decaying phase of the XRT afterglow might be
the low energy tail of the prompt emission.
The second break in the afterglow light curve occurs about $27000$~s
after the burst. 
The spectral energy distribution before and after the second break 
does not change and it can be tentatively interpreted as a jet break
or the end of a delayed or continuous energy injection phase. 

\end{abstract}


\section{Introduction}
\label{section:introduction}

The Swift Gamma-Ray Burst Explorer (Gehrels et al. 2004), successfully launched 
on 2004 November 20, is  
dedicated to the discovery and study of gamma-ray bursts and their
X-ray and optical afterglows.
Its fast autonomous pointing capability, compared  to previous satellites,
allows Swift to repoint towards GRB sources approximately 100 s after the burst
detection by the Burst Alert Telescope (BAT) and to study, for the
first time, the early phases of the afterglow evolution.
Moreover, the very broad energy band allows for a simultaneous study
of the afterglow in the optical, soft and hard X-ray bands.

The Burst Alert Telescope (Barthelmy et al. 2005a) on board Swift detected
and located GRB\,050319 at 09:31:18.44 UT (Krimm et al. 2005a).
A re-analysis of the BAT data showed that the GRB050319 onset
was $\sim$ 135 s before the trigger time given in Krimm et al. (2005a). 
The GRB light curve was characterized by several peaks, spaced out in time.
The Swift spacecraft was slewing
at the epoch of the GRB onset and the BAT trigger was
switched-off. The GRB was recognized only during the last peak, 50-60 s
after the end of the slew, activating the Swift GRB follow-up sequence.
The measure of the entire burst duration yields a $T_{90}$ = 149.6 $\pm$ 0.7 s.
The time-averaged energy distribution was  modeled
by a simple power law with a photon index $\Gamma$ = 2.1 $\pm$ 0.2
(90\% confidence level) in the 20-150 keV energy range.
The burst fluence over the $T_{90}$ interval in the 15-350 keV band 
was 1.6 $\times$ 10$^{-6}$ erg cm$^{-2}$.
At the known redshift of 3.24, the isotropic equivalent $\gamma$--ray energy is
3.7 $\times$ 10$^{52}\,$erg \footnote{We used
$\Omega_{\rm M}$=0.27, $\Omega_{\Lambda}$=0.73, $H_0$=71\,km\,s$^{-1}$\,Mpc$^{-1}$}.
Given the soft spectrum of this burst ($\Gamma\sim2$),
the bolometric correction is small.

The spacecraft executed an immediate slew and promptly began observing 
with its narrow field instruments to monitor the afterglow.
The X-Ray Telescope (XRT; Burrows et al. 2005) and the Ultraviolet and Optical 
Telescope  (UVOT; Roming et al. 2004, 2005) were on target at 09:32:45.53 UT, 90 s after
the BAT trigger, i.e. $\sim 220$ s after the burst onset.
UVOT revealed a new source inside the BAT error circle at
RA$_{\rm J2000}$ = 10h16m47s.76(3), 
Dec$_{\rm J2000}$ = +43$^{\circ}$ 32$\arcmin$ 54$\farcs$9(5) (Boyd et al. 2005).
The source intensity faded, during a time interval of 17000 s, from 17.5 to 20.6
mag and from 18.8 to 21.2 mag in the $V$ and $B$ band, respectively. 
No signal was detected in the $U$ band with an upper limit of 19 mag.
A full description of UVOT results is presented in Mason et al. (2005).

Ground-based follow-up observations were performed in different wavelength bands.
ROTSE-IIIb detected a 16th mag fading source at RA$_{\rm J2000}$ = 10h16m47s.9, 
Dec$_{\rm J2000}$ = +43$^{\circ}$32$\arcmin$54$\farcs$5, 
27 s after the trigger. The intensity decreased to the 18th mag after 940 seconds
(Rykoff et al. 2005)
following a power law ($F(t,\nu) \propto t^{-\alpha} \nu^{-\beta}$) 
with $\alpha$ = 0.59 $\pm$ 0.05 (Quimby et al. 2005). 
The intensity decay of the optical afterglow was
confirmed by Kiso 1.05-m,  Lulin 1-m, ART 14-inch and Mt. Maidanak 1.5-m
telescopes (Yoshioka et al. 2005; Torii 2005; Sharapov et al. 2005a,b).  
Wo\`zniak et al. (2005) presented extensive early-time photometry starting
25.4\,s after the trigger. From their detailed light curve, one of the
best sampled available to date, they infer a break time at about 450\,s
after the trigger with pre- and post-break slopes 0.37 and 0.91.
Spectra of the GRB\,050319 afterglow obtained by the
Nordic Optical Telescope revealed several absorption features, including
strong Ly$\alpha$, OI+SiII, SiIV and CIV. Their detection implies a
redshift $z=3.24$ (Fynbo et al. 2005).
GRB\,050319 was also detected in the near infrared by the NICMOS camera
in the PRL 1.2m telescope to a $J$-band mag of about 13 
(George et al. 2005) 6.2 hours after the burst.
Radio 8.5 GHz observations with the Very Large Array did not detect a radio
source at the position of the optical afterglow; the derived  2$\sigma$
upper limit was 70 $\mu$Jy (Soderberg 2005a,b). 

In the following,  we report the results on the GRB afterglow observed by
the Swift XRT. The details of the X-ray observations and the data reduction
are described in  Section 2; temporal and spectral analysis
results are reported in section 3 and discussed in section 4 where  
we draw our conclusions.

\section{Observations and Data Reduction}
The Swift XRT is 
designed to perform automated observations of newly discovered
bursts in the X-ray energy band 0.2-10 keV.  
Four different read-out modes have been implemented, each dependent on the 
count rate of the observed
sky region. The transition between two  modes is automatically performed  on board
(see Hill et al. 2004 for a detailed description on XRT
observing modes). 

XRT was on target 90 s after the BAT trigger and
it observed GRB\,050319 for 17 consecutive orbits for a total elapsed time of 
105678 s.  
Moreover, GRB\,050319 was further observed several times up to 28 days later. 
The observation log is
presented in Table 1.

During the first observation, XRT observed in windowed timing (WT) 
and photon counting  (PC) mode, in the later observations only PC mode
was used. In the WT mode only the central $8\arcmin$ of the field of view
is read out, providing one dimensional imaging and full spectral
capability with a time resolution of 1.8 ms. The PC mode provides,
instead, full spatial and spectral resolution with a timing resolution of
2.5 seconds.

\begin{deluxetable}{cccccc}
\tablecolumns{6}
\tabletypesize{\tiny}
\tablecaption{XRT Observation log of GRB\,050319}
\tablewidth{0pt}
\tablehead{
\colhead{Observation \#} &
\colhead{Start Time} &
\colhead{End Time} &
\colhead{Start Time since GRB onset} &
\colhead{WT Exposure} &
\colhead{PC Exposure}\\
\colhead{} &
\colhead{(yyyy-mm-dd hh:mm:ss) UT} &
\colhead{(yyyy-mm-dd hh:mm:ss) UT} &
\colhead{(s)} &
\colhead{(s)} &
\colhead{(s)} 
}
\startdata
1  & 2005-03-19 09:33:02   & 2005-03-20 14:54:20 & 222.7        & 9443 & 12985  \\
2  & 2005-03-24 01:52:04   & 2005-03-24 05:34:44 & 410120.0     & --   & 986  \\
3  & 2005-03-26 10:16:04   & 2005-03-28 10:50:53 & 607651.0     & --   & 8486  \\
4  & 2005-04-07 03:40:55   & 2005-04-07 19:48:45 & 1620708.0    & --   & 2844  \\
5  & 2005-04-08 03:48:41   & 2005-04-08 16:45:57 & 1707573.0    & --   & 1101  \\
6  & 2005-04-09 16:45:30   & 2005-04-11 05:57:21 & 1840723.0    & --   & 3639  \\
7  & 2005-04-13 01:28:15   & 2005-04-13 22:16:58 & 2131148.0    & --   & 2397  \\
8  & 2005-04-14 09:26:38   & 2005-04-14 14:30:58 & 2246250.0    & --   & 754   \\
9  & 2005-04-15 06:06:40   & 2005-04-15 21:00:59 & 2320473.0    & --   & 3371  \\
10 & 2005-04-16 01:20:12   & 2005-04-16 14:11:23 & 2389865.0    & --   & 3891  \\
\enddata
\label{tab2}
\end{deluxetable}

XRT data were first processed by the Swift Data Center at NASA/GSFC
to produce calibrated event lists (level 1 data products). 
They were therefore filtered and screened using the XRTDAS (v.1.2) 
software package to produce cleaned photon list files. Only observing time 
intervals with a CCD temperature below -50 degrees Celsius were used.
Further non-standard selections were performed to remove time
intervals with high background rate caused by either dark current
or by the bright Earth limb.
The total exposure after all the cleaning procedures were 9443 s and 40454 s
for data accumulated in WT and PC mode, respectively.
Hot and flickering pixels were further removed with an ad hoc
procedure.

For the spectral analysis we used a 0-4 grade selection for data in PC mode.
Such a selection provides the best combination of spectral resolution
and detection efficiency. 
Ancillary response files were generated for each spectrum through the
standard {\it xrtmkarf} task (v0.4.13) with an input mirror file 
obtained using experimental gold reflectivity coefficients 
derived by Owens et al. 1996. 
Response file {\it swxpc0to4\_20010101v007.rmf} 
has been used to fit PC spectra.

For the timing analysis we selected events with 0-12 and 0-2 grades
for the PC and WT data, respectively. 
This selection maximises the light curve statistics.
Hereafter, errors are reported with a 90\% single parameter
confidence level.

The XRT times are referred to the GRB050319 onset 
$T=$2005 Mar 19.39517 UT (2005 Mar 19, 09:29:02.70 UT). 

\section{Data Analysis}

\subsection{Spatial analysis}
Fig. 1 shows the XRT image accumulated in PC mode with a 0.2-10 keV energy
selection.
The GRB afterglow position derived with {\it xrtcentroid} (v0.27) 
is  RA$_{\rm J2000}$ = 10h16m48s.1, Dec$_{\rm J2000}$ = +43$^{\circ}$$32\arcmin$52$\farcs$4. 
The GRB  position uncertainty is 6$\arcsec$  (90\% confidence level). 
This uncertainty is mostly due to the preliminary calibration 
of the boresight accuracy and to the residual misalignment
between the XRT optical axis and the star trackers. The afterglow candidate
position is consistent with the UVOT new source position (Boyd et al. 2005). 

In the same figure we plot the XRT error box circle together with the BAT error box and
the optical counterpart coordinates derived by UVOT.
The XRT derived coordinates are $30$\farcs$5$ from the BAT ones (Krimm et al. 2005a) 
and 4$\farcs$9 from the optical counterpart (Boyd et al. 2005).
 
\subsection{Timing analysis}  
During the first orbit the spacecraft was pointing 
5$\arcmin$.7 away from the BAT GRB centroid computed by
the on board software, causing the GRB\,050319 centroid to fall 4 pixels outside the 
XRT WT window.
WT data from this orbit were extracted from a rectangular region 16 pixels wide at the 
boundary of the image strip. 
These counts were then corrected taking into account the Point Spread 
Function (PSF) fraction of the selected region. 
The correction factor was evaluated by 
weighting the energy dependent PSF by the counts extracted
from the selected region.
The selected extraction region amounts to about 
14\% of the XRT PSF.
This correction allowed us to coherently include the first orbit WT data 
in the light curve together with PC data. The spacecraft pointing was
corrected in the following orbits and the GRB image was centered in
the WT window. All the following WT data were extracted in a rectangular region 40 pixels wide 
along the image strip that includes about 98\% of the PSF. 

The intensity of the source during the first four orbits 
is high enough to cause a pileup in the PC frames. 
In order to correct for the pileup, we
extracted counts from an annular region with an outer radius of 30 pixels (70$\farcs$8)
and an inner radius of 6 pixels (14$\farcs$16). These values were evaluated by comparing
the analytical PSF with the profile extracted in the first 30 seconds 
of observation when the intensity has the highest value. 
The extracted counts were corrected for the fraction of the PSF of the selected region. 
Again, the correction factor was evaluated by 
weighting the energy-dependent PSF by the counts extracted
from the annular region.
Such a region amounts to about 30\% of the PSF. 
In the following 13 orbits data were extracted from the entire
circular region of 30 pixels radius to have the maximum 
available statistics, particulary important in the last part of the afterglow decay light curve.

The background level for the WT light curve was extracted from a rectangular 
region 80 pixels wide, far from the source and affected by minimal
contamination from other sources in the field.
The background level for the PC data was extracted in an annular region with 
an inner radius of 50 pixels and an outer radius of 100 pixels centered 
on the source position  and free from other sources contribution. 
The background values, normalized to the source extraction regions, 
were 2.1 $\times$ 10$^{-3}$ counts\,s$^{-1}$ and 5.0 $\times$ 10$^{-2}$ 
counts\,s$^{-1}$ for PC and WT, respectively.
The background level did not vary with time and the same 
average values were then used over the whole light curve. 
In the first GRB observation, WT and PC data were binned in order to 
have a constant signal to noise 
ratio of 4.5.
In the second and third observations (Table 1) the source is barely detectable. 
We extracted only one time bin for each of them with a significance of 2.0 and 2.6
standard deviations, respectively. 
From observation 4 to 10 we derived only an upper limit. 
Fig. 2 shows the background-subtracted light curve in the 0.2-10 keV energy band. 
The source is clearly fading with time. 

We fit the X--ray light curve, using only data from the first observation. 
The light curve decay is not consistent with a single power law 
($\chi_{\rm red}^{2} = $ 2.0, with 78 d.o.f.). 
A broken power law 
($F(t)=K~t^{-\alpha_{\rm A}}$ for $t < T_{\rm break,1}$ and 
$F(t)=K~T_{\rm break,1}^{\alpha_{\rm B}-\alpha_{\rm A}}~t^{-\alpha_{\rm B}}$ 
for $t \ge T_{\rm break,1}$) 
improves the fit, giving a break 
at 370 $\pm$ 18 s from the GRB onset and a resulting $\chi_{\rm red}^{2} = $
1.3 (76  d.o.f.). 
However, the residuals show a systematic trend with a
negative slope after the break time. 
Adding a second break to the model 
($F(t)=K~t^{-\alpha_{\rm A}}$ for $t < T_{\rm break,1}$;
$F(t)=K~T_{\rm break,1}^{\alpha_{\rm B}-\alpha_{\rm A}}~t^{-\alpha_{\rm B}}$  
for $ T_{\rm break,1} \le t < T_{\rm break,2}$ and
$F(t)=K~T_{\rm break,1}^{\alpha_{\rm B}-\alpha_{\rm A}}~T_{\rm break,2}^{\alpha_{\rm C}-\alpha_{\rm B}}~t^{-\alpha_{\rm C}}$ 
for $t \ge T_{\rm break,2}$) 
the fit improves to a $\chi_{\rm red}^{2} = $  0.75
(74 d.o.f.) and the F$-$test yields a chance probability of 
6.0 $\times$ 10$^{-10}$. This last model reveals the presence of a second  break
at (2.7 $\pm$ 0.7)  $\times$ 10$^4$ s.
Table 2 shows the best fit results obtained with the three models.
In Fig. 2 we also plot the best fit model obtained with the broken power law
with two breaks.
Moreover, Fig. 2 shows that the extrapolation of the best fit model
is consistent with the detections obtained in the second and third observations
and with the 3 $\sigma$ upper limit of 4.43 $\times$ 10$^{-14}$\,erg\,cm$^{-2}$\,s$^{-1}$ 
derived from observations 4-10.
We also tried to fit the light curve with a power law or a single break broken power law 
allowing the reference time $t_0$ to be a free parameter.
The $\chi^2$ did not improve, and the best fit $t_0$ was consistent with zero
i.e. with the GRB onset) for the power law ($t_0$ = 800 $\pm$ 1000 s) 
and with a negative value (i.e. a time before the GRB onset) 
for the broken power law ($t_0$ = -74 $\pm$ 26 s).

We checked that our results are not affected by the PSF correction we applied 
on both PC and WT data: we 
produced a PC light curve by selecting counts only in 
the annular region of 6 and 30 pixels radii for all the first 
observation.
Again the best fit model is a broken power law with two breaks;  
the best fit parameters are consistent 
within the errors with those reported in Table 2. In this case, however, the significance 
level of the second break is lower since 
this extraction region contains only 30\% of the source counts.

\begin{deluxetable}{cccc}
\tablecolumns{4}
\tabletypesize{\normalsize}
\tablecaption{GRB\,050319 light curve best fit parameters}
\tablewidth{0pt}
\tablehead{
\colhead{Parameter} &
\colhead{Single power law}&
\colhead{Broken power law}&
\colhead{Double broken power law} 
}
\startdata
$\alpha_{\rm A}$         & 0.67 $\pm$ 0.02              & 5.53 $\pm$ 0.66      & 5.53 $\pm$ 0.67      \\
$T_{\rm break,1}$ (s)    &  --           & 370 $\pm$ 18 & 384 $\pm$ 22 \\
$\alpha_{\rm B}$         &  --           & 0.65 $\pm$ 0.0 & 0.54 $\pm$ 0.04      \\
$T_{\rm break,2}$ (s)    &  --           &  --          & (2.60 $\pm$ 0.70) $\times$ 10$^4$     \\
$\alpha_{\rm C}$         &  --           &  --          & 1.14 $\pm$ 0.2       \\
$\chi_{\rm red}^{2}$ (d.o.f.)   &  2.02 (78)            &  1.2 (76)     & 0.75 (74)     \\ 
\enddata
\label{tab2vvv}
\tablecomments{$\alpha_{\rm A}$, $\alpha_{\rm B}$ and $\alpha_{\rm C}$ are the decay slopes for
the distinct phases of the afterglow (see section 3.3). $T_{\rm break,1}$ and $T_{\rm break,2}$
are the epochs of the decay slope discontinuity, measured from the GRB onset.}
\end{deluxetable}

\subsection{Spectral analysis}  
The presence of two breaks in the decay light curve of GRB\,050319 suggests 
looking for spectral variations across the breaks.
We therefore extracted three spectra from the PC data of the first observation,
corresponding to the three time intervals delimited by the two breaks.
Hereafter  we call them A, B and C.
Spectrum A was extracted from the same annular region used for the 
timing analysis of the first four orbits. In the second interval pileup is still present but the 
lower intensity of the source allows us to reduce the inner radius of the 
annular region to 4 pixels. Thus, spectrum B is extracted from an annular region
including 44\% of the PSF. Spectrum C was extracted from the entire 
circular region of 30 pixels radius. 
We ignored  energy channels below 0.2 keV and above 10 keV and rebinned  
the spectra to at least 20 counts per bin to allow the use of $\chi^2$ statistics. 
The background spectrum was extracted 
from the same region as for the timing analysis.

The spectrum of each interval was fitted with an
absorbed power law with the absorption column density fixed to the 
Galactic value of 1.13 $\times$ 10$^{20}$ cm$^{-2}$ (Dickey \& Lockman 1990). 
This model gives a good description of data for all the three spectra. 
A blackbody model does not reproduce the energy distribution of the 
three phase interval spectra ($\chi_{\rm red}^{2} >$ 5). 
The photon index of spectrum A is significantly softer than the other two.
The XRT spectra for the three intervals and the spectral residuals are shown in Fig. 3.
Table 3 shows best fit parameters.

We checked for intrinsic absorption in the host galaxy by adding
a redshifted observation component (ZWABS model in XSPEC v11.3.1)
with redshift fixed to 3.24 (Fynbo et al. 2005). Only spectra B and C
had enough counts to permit this test.
The fits gave  a value for the additional column density  of 
(0.38 $\pm$ 0.22) $\times$ 10$^{22}$ cm$^{-2}$ and (0.37 $\pm$ 0.35) $\times$ 10$^{22}$ cm$^{-2}$
for B and C, respectively. However, the improvement in the fits is not 
significant.

The soft photon index in spectrum A is in agreement with the results
from the hardness ratio analysis: the interval A yields a (3-10 keV)/(0.2-3 keV) of 0.075 $\pm$ 0.02, 
the interval B$+$C gives a value of 0.14 $\pm$ 0.01.

\begin{deluxetable}{cccc}
\tablecolumns{4}
\tabletypesize{\tiny}
\tablecaption{GRB\,050319 Spectral fit results}
\tablewidth{0pt}
\tablehead{
\colhead{Parameter} &
\colhead{A}&
\colhead{B}&
\colhead{C} 
}
\startdata
Galactic $N_{\rm H}$ (10$^{20}$ cm$^{-2})$                 & 1.13                  & 1.13          &  1.13 \\
$\Gamma$        &  2.60 $\pm$ 0.22             & 1.69 $\pm$ 0.06      & 1.8 $\pm$ 0.08       \\
$N$ (photons keV$^{-1}$  cm$^{-2}$ s$^{-1}$ at 1 keV)  &  (1.7 $\pm$ 0.3) $\times$ 10$^{-2}$           & (1.53 $\pm$ 0.09)  $\times$ 10$^{-3}$     & (3.96 $\pm$ 0.3)  $\times$ 10$^{-4}$      \\
Flux$_{ 0.2-10 \rm keV}$ (erg cm$^{-2}$ s$^{-1}$) & (9.2 $\pm$ 0.8) $\times$ 10$^{-11}$      & (1.1 $\pm$ 0.04) $\times$ 10$^{-11}$       & (2.68 $\pm$ 0.13) $\times$ 10$^{-12}$      \\
$L_{ 0.2-10 \rm keV}$ (erg s$^{-1}$) & 1.9 $\times$ 10$^{49}$              &  8.3 $\times$ 10$^{47}$              &    2.3 $\times$ 10$^{47}$           \\
$\chi_{\rm red}^{2}$ (d.o.f.) & 1.1 (9)       & 1.4 (42)      & 0.8 (18)      \\
\enddata
\label{tab3vvv}
\tablecomments{$\Gamma$ is the photon index. $L$ is the isotropic luminosity calculated for a redshift $z = 3.24$. 
Fluxes and luminosities reported here are averaged over long time intervals. 
Accurate instantaneous values for the flux should be derived from Fig. 2.}
\end{deluxetable}


\begin{figure}[ht]
\vspace{1.0cm}
    \figurenum{1}
    \epsscale{0.8}
\centerline{\psfig{figure=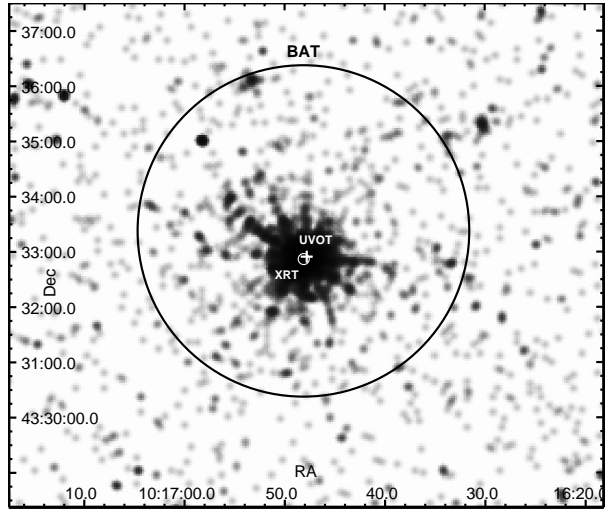,width=8cm,angle=0}}
    \caption{Photon counting mode XRT image with the XRT and BAT error circles 
and the UVOT position. 
      }
\end{figure}

\begin{figure*}[ht]
    \figurenum{2}
    \epsscale{0.8}
\centerline{\psfig{figure=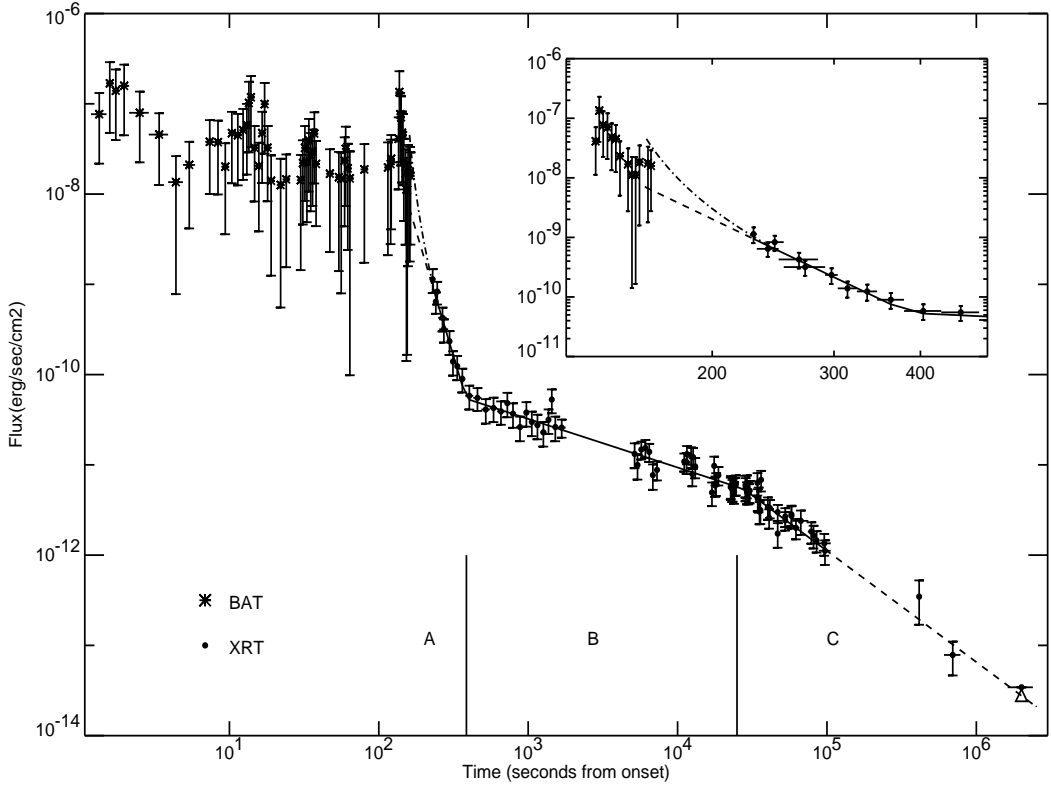,width=12cm,height=16cm,angle=+90}}
    \caption{XRT light curve decay of GRB\,050319. The XRT count rate (0.2-10 keV)
was converted into flux units by applying a conversion factor derived 
from the spectral analysis of each correspondent phase (see section 3.3). The solid line
represents the best fit model, the dashed line is the extrapolation of 
this model prior to the first XRT observation. The dot-dashed line
represents the extrapolation back to the end of the prompt emission of
the best fit obtained using the double broken power law model 
with times referred
to the peak time of last spike of the prompt emission ($T$= 137 s
after the burst onset). The last point after 10$^6$ s is a 3 $\sigma$ upper limit. 
The gaps in the XRT light curve in the first observation are due to blind observing periods during
the spacecraft orbit. 
The BAT light curve was extrapolated into the XRT energy band 
by converting the BAT count rate with the factor derived from the BAT 
spectral parameters. The gaps in the BAT light curve corresponds to time intervals where
the GRB count rate level was consistent with zero. The inset shows the time interval including
the last peak of the GRB and the phase A of the afterglow.
      }
\vspace{1.0cm}
\end{figure*}

\begin{figure*}[ht]
    \figurenum{3}
    \epsscale{0.8}
\centerline{\psfig{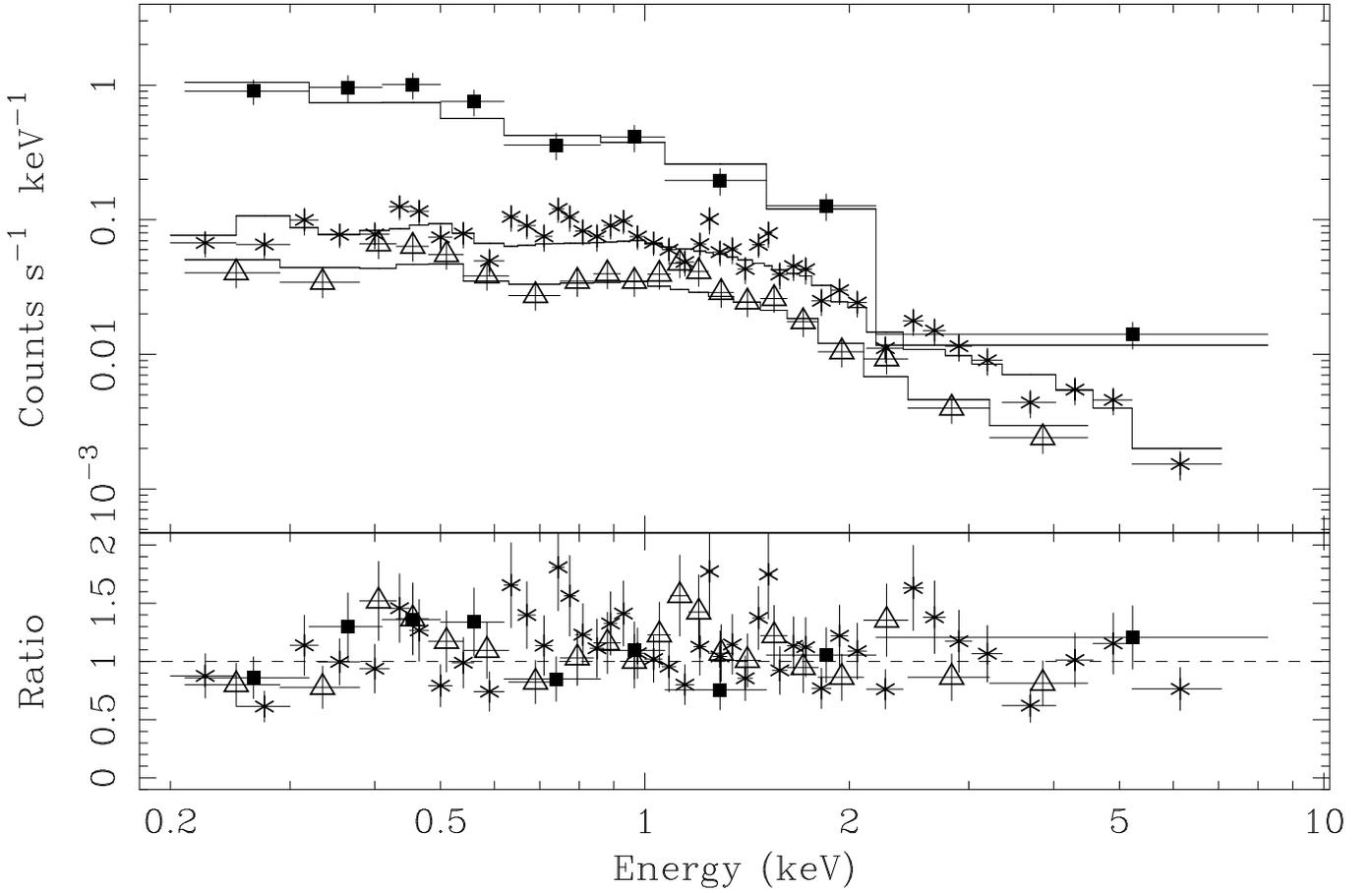}}
    \caption{XRT 0.2-10 keV energy spectrum of GRB\,050319 afterglow, 
with the
      best-fit absorbed power law model. The spectra of 
      the three intervals (A: squares, B: stars, C: triangles) are plotted in the upper panel. The 
      residuals are plotted in the lower panel.}
\end{figure*}

\section{Discussion}

XRT monitored GRB\,050319 X-ray emission from $\sim$ 220 s after the burst onset 
up to 28 days.
The afterglow light curve is characterized by an initial steep decay 
($\alpha_{\rm A} \sim 5.5$) followed by a less rapid decline. 
Similar steep declines followed by a flattening have been 
observed in other GRB afterglows promptly detected by Swift 
thanks to its unprecedentedly fast repointing capability: GRB\,050117a
(Hill et al. 2005, in preparation),
GRB\,050126, GRB\,050219a (Tagliaferri et al. 2005), and
GRB\,050315.

What is remarkable for the GRB\,050319 X-ray afterglow is that 
the intensity decay cannot be modeled either by a simple power law 
or by a broken power law with only one break. 
The GRB050319 light curve of the first $10^5$ s after the trigger
shows evidence of three different phases (A, B and C according to section 3.3), 
each of them characterized by a distinct decay slope (see Fig. 2 and Table 2). 
We also found that the extrapolation of phase C decay
is consistent with the emission detected 8 days after the burst 
and with the flux upper limit measured after 28 days.
 
The spectral analysis shows that in phase A
the spectral energy distribution is significantly softer 
($\beta = \Gamma -1 = 1.6 \pm 0.22$) 
than in phases B and C, where the two $\beta$ are consistent; 
their weighted average is $\beta = 0.73 \pm 0.05$. 
This might indicate a different emission process acting during the initial 
phase of GRB\,050319 afterglow. 

Fig. 2 reports the extrapolation to the 0.2-10 keV band of the 
BAT prompt light curve. Fluxes were evaluated adopting a conversion
factor computed by extrapolating the BAT spectral best fit model
to the XRT energy band. The first steep phase of the afterglow 
light curve, extrapolated back to the trigger epoch, 
is consistent with the BAT fluxes.
Moreover, the photon indices of the last peak of the prompt emission 
(2.2 $\pm$ 0.2; Krimm et al. 2005b)
and of the phase A X--ray afterglow (2.6 $\pm$0.2; Table 3) are
marginaly consistent within the errors.
This indicates that the early, steep and soft X--ray afterglow
might be the low energy tail of the last peak of the GRB emission.
Thanks to the XRT better sensitivity and lower energy band 
we can in principle observe prompt emission in X--rays longer 
than in $\gamma$-rays. 
If our hypothesis is correct, the data of Phase A should be referred 
to the peak time of last spike of the prompt emission ($T$= 137 s 
after the burst onset). In fact, the tails of each previous 
spike would largely have decayed at the time of
the Phase A and only the last spike would be relevant.
In such a way, we obtain a decay slope for segment A of 2.9$\pm$0.5 and 
the model extrapolation (see the inset in Fig. 2) back to the end of the 
prompt emission is marginaly consistent
with the end of the BAT data. A more detailed analysis of the evidences
recentely provided by Swift that early time X-ray fluxes can represent 
the lower energy tail of the prompt emission is presented in Barthelmy et 
al. 2005b. The authors use GRB050319  as one of the best examples.
In this scenario the first break might be due to the emergence of the
afterglow after fading of the GRB last peak emission. 

The most straightforward interpretation of the second temporal break 
is a jet break. Within this scenario, a steepening of the observed emission
occurs when the relativistic beaming angle of the emitted
radiation becomes larger than the geometrical opening angle of the jet 
because of the deceleration caused by the interaction with 
the external medium.
If we assume that the external medium is a uniform interstellar medium
with particle density $n$,
then we can assume that emission in phases B and C is due to synchrotron 
radiation from shock-accelerated electrons in the slow cooling regime
with $\nu > \nu_c$, where  $\nu_c$ is the cooling frequency.
The latter is a standard condition in the X-ray band. 
The electron energy distribution index $p$ determines 
both the light curve slope 
and the spectral index before and after the jet break 
(Rhoads 1999). 
The temporal index before the jet break, when the spherical
expanding shell model can be used, is expected to be
$\alpha_1$ = $(3p-2)/4$, while after the jet break 
$\alpha_2$ = $p$ is foreseen. The spectral index $\beta=p/2$ is
expected to remain unchanged before and after the break.
This model is consistent with the second break of the GRB\,050319 
afterglow having $\beta\sim 0.7$, $\alpha_1=\alpha_{\rm B}\sim 0.5$ 
and $\alpha_2=\alpha_{\rm C}\sim 1.2$ for $p=1.4$.   
Such a low value of the electron distribution index is
not common in late afterglow fits, but could be generated
by the mechanism discussed by Bykov \& M\'{e}sz\'{a}ros (1996). 
According to this interpretation, the second break in the
light curve would imply a jet opening angle of 
$\theta_0 = 2.3^{\circ} \left(t_b / 26000 {\rm s}\right)^{3/8} 
n_0^{1/8}\left(\eta_{\gamma}/ 0.2 \right)^{1/8}$, 
where $n_0=n / 1$\,cm$^{-3}$, $t_b$ is the break time, $\eta_{\gamma}$ is the 
conversion efficiency of internal energy to $\gamma$-ray radiation
and we used the values 3.24 and 3.7$\times 10^{52}$~erg
for the redshift and the isotropic equivalent emitted energy
in $\gamma$-rays respectively (Sari et al. 1999).
This value of the opening angle is on the low-end tail of the
observed distribution of jet opening angles presented by
Bloom, Frail and Kulkarni (2003). 
Moreover, the time of this break is earlier than
typical jet breaks which are seen at t $> 10^5$\,s. 
Also, the initial and final slopes are both flatter 
than previous jet break observations. 

A possibility to avoid low $p$ values and
explain both the second break in the afterglow 
and the unusually flat decay slope during phase B 
is based on refreshed shocks (Sari \& M\'{e}sz\'{a}ros 2000)
or continuous energy injection from a Poynting flux
dominated flow (Zhang \& M\'{e}sz\'{a}ros 2002).
Note that the phase C decay slope and spectral index
are consistent with $\alpha_2=3(p-1)/4$ and $\beta=(p-1)/2$
for $p \sim 2.5$. This is what expected for fireball
expansion in a uniform ISM when $\nu_m < \nu_X < \nu_c$
(here $\nu_X$ represents the typical X--ray frequency and
$\nu_m$ is the characteristic frequency of synchrotron 
radiation from shock-accelerated electrons). 
Within this scenario, delayed energy injection due to
the catching up of slower shells
with the bulk of the expanding flow or 
long lasting emission from the central engine 
in the form of a Poynting flux dominated outflow 
could be responsible for the flatter decay of the afterglow 
in phase B. In this case a transition to the standard 
afterglow evolution (i.e. a break) with no remarkable 
spectral changes is expected when the
addiditonal energy supply ends.
Consequences of a continuously injecting central engine
like a highly magnetized millisecond pulsar  
on GRB afterglow emission have been investigated
by Zhang \& M\'{e}sz\'{a}ros (2001). According to their
analysis, for an injection law of the form $\dot{E}(t) \propto t^{-q}$,
the injection would influence the fireball evolution if $q<1$.
A decay slope $\alpha_{\rm inj}=(1+q/2)\beta+q-1$ is expected 
for the afterglow light curve until injection stops. For 
GRB\,050319, being $\beta\sim0.7$ and $\alpha_{\rm inj}=\alpha_{\rm B}\sim0.5$, 
we can derive $q=0.5$--$0.6$. This value would be explained
by a central long--lived black hole--torus system with reduced
activity.
This interpretation would predict also a cooling break 
at later times with $\alpha_2$ and $\beta$ becoming steeper 
by 1/4 and 1/2 respectively when the cooling frequency $\nu_c$
crosses the X--ray band. According to Panaitescu \& Kumar (2000)
the non detection of such a further
break in the X--ray light curve up to about $10^6$ seconds
after the trigger could be explained by a low density medium
($n < 10^{-3}$\,cm$^{-3}$) and a low magnetic field
in the post-shock fluid 
($\epsilon_{\rm B} < 5 \times 10^{-3}$).

A rich optical dataset is available for this burst, even if part of the 
data are badly calibrated.
For an extended discussion about them we refer to Mason et al. (2005)
and to Wo\'zniak et al. (2005).
About the latter paper, we only want to stress that, as the authors
were unaware of the true GRB onset time $\sim135$\,s before the trigger, 
their break time and temporal slopes are referred to the trigger time.
A very simple time shift of their data gives a break time at 614.2$\pm$0.6\,s 
(not consistent with any of the breaks in the X--ray light curve)
and slopes before and after the break equal to 0.71$\pm$0.06 and 1.00$\pm$0.07
respectively, with a $\chi^2_r$ of 3.3. 
The data are also consistent with a single power law decay with slope 0.84$\pm$0.03 
($\chi^2_r=4$). With this new reference time,
the presence of a break may even no longer be required: the introduction
of a break time improves the fit, but the significance of the improvement
is not compelling according to an F-test (chance probability $3\times10^{-2}$).
The non correspondance of any of this slopes to what seen
in X--rays is clearly challenging standard interpretations.

\section{Acknowledgments}
This work is supported at INAF by funding from ASI on grant number I/R/039/04,
at Penn State by NASA contract NASS5-00136 and at the University of Leicester by
 the Particle
Physics and Astronomy Research Council on grant numbers PPA/G/S/00524 and PPA/Z/
S00507. We
gratefully acknowledge the contribution of dozens of members of the XRT team at
OAB, PSU, UL, GSFC,
ASDC and our subcontractors, who helped make this instrument possible.

\end{document}